\documentclass[journal]{IEEEtran}

\ifCLASSINFOpdf
\else
   \usepackage[dvips]{graphicx}
\fi
\usepackage{url}
\usepackage{cite}
\usepackage{amsmath,amssymb,amsfonts}
\usepackage{algorithmic}
\usepackage{textcomp}
\usepackage{multirow}
\usepackage{balance}
\usepackage[table,xcdraw]{xcolor}
\usepackage{graphicx}
\newcommand{\blue}[1]{\textcolor{black}{#1}}

\begin{document}

\title{XLSR-Mamba: A Dual-Column Bidirectional\\ State Space Model for Spoofing Attack Detection}


\author{Yang Xiao and Rohan Kumar Das,~\IEEEmembership{Senior Member, IEEE}
\thanks{Yang Xiao and Rohan Kumar Das are with Fortemedia, Singapore 138589 (email: \{xiaoyang, rohankd\}@fortemedia.com). }}

\markboth{Journal of \LaTeX\ Class Files, Vol. 14, No. 8, August 2015}
{Shell \MakeLowercase{\textit{et al.}}: Bare Demo of IEEEtran.cls for IEEE Journals}
\maketitle

\begin{abstract}
Transformers and their variants have achieved great success in speech processing. However, their multi-head self-attention mechanism is computationally expensive. Therefore, one novel selective state space model, Mamba, has been proposed as an alternative. Building on its success in automatic speech recognition, we apply Mamba for spoofing attack detection. Mamba is well-suited for this task as it can capture the artifacts in spoofed speech signals by handling long-length sequences. However, Mamba's performance may suffer when it is trained with limited labeled data. To mitigate this, we propose combining a new structure of Mamba based on a dual-column architecture with self-supervised learning, using the pre-trained wav2vec 2.0 model. The experiments show that our proposed approach achieves competitive results and faster inference on the ASVspoof 2021 LA and DF datasets, and on the more challenging In-the-Wild dataset, it emerges as the strongest candidate for spoofing attack detection. The code has been publicly released\footnote{https://github.com/swagshaw/XLSR-Mamba}.
\end{abstract}

\vspace{-2mm}

\IEEEpeerreviewmaketitle
\vspace{-2mm}
\section{Introduction}
In recent years, the rapid advancement of deep learning has led to significant improvements in speech generation technologies such as text-to-speech (TTS) synthesis~\cite{blizzard2023} and voice conversion (VC)~\cite{VCC2020}. While these technologies offer many benefits, they also introduce serious risks when misused to produce fake speech for a target speaker. The latest TTS and VC systems are able to produce very high-quality natural-sounding speech, posing a threat to security and trust in voice-based systems. As a result, automatic speaker verification (ASV) systems are increasingly vulnerable to attacks derived using TTS and VC~\cite{survey1,asv,survey2}. 
Developing anti-spoofing solutions is critical to counter these attacks. The ASVspoof challenge series has gained widespread attention by offering standardized datasets, evaluation rules, and performance metrics for the anti-spoofing research community~\cite{ASVspoof_journal}. 

Previous works in anti-spoofing advocated long-term features to capture the relevant artifacts for spoofing attack detection~\cite{PatelINTERSPEECH2015,CQCC_CSL,rkd_is2019,rkd_ASRU2019}. Adding value to these works, recent research shows that the cues discriminating spoofed speech from bonafide ones can be well apprehended by local and global features~\cite{rawnet2,rawformer}. Local features may reveal unnatural intonation or stress, while global features often exhibit irregular rhythms or flat emotional tones. Combining information from both levels enhances detection performance. 

End-to-end models optimize directly on the raw audio without relying on handcrafted features and have shown great potential. Therefore, much of the current research is focused on such end-to-end models. For instance, RawNet2 applies time-domain convolution on the raw audio effectively learning the features~\cite{rawnet2}. Attention mechanisms have also been integrated into graph neural networks to better capture important information across time and frequency. Models like AASIST utilize graph attention networks to process complex local and global features~\cite{aasist}. Similarly, Rawformer combines convolutional layers with transformer structures, using self-attention to extract important cues for anti-spoofing~\cite{rawformer}. 


Self-supervised learning (SSL)~\cite{xlsaasist} has gained attention for creating pre-trained models that generalize well with limited labeled data. Several SSL speech models, including auto-regressive predictive coding, wav2vec~\cite{wav2vec}, and HuBERT~\cite{hubert}, have demonstrated promising results in various speech-processing tasks. A recent anti-spoofing model combined the SSL model wav2vec 2.0 XLS-R (0.3B)~\cite{wav2vec2,xls} with the transformer-based conformer architecture~\cite{asvconformer, conformer}, achieving state-of-the-art (SOTA) results in ASVspoof 2021~\cite{asvspoof2021}. This success is linked to the multi-head self-attention (MHSA)~\cite{vaswani2017attention} mechanism’s strong sequence modeling capabilities. 
However, empirical evidence\blue{~\cite{mambaspeech}} shows that MHSA is less effective at selectively capturing relevant information for modeling long-range relationships. Therefore, we are interested in proposing a more effective method for capturing long-range feature information in speech signals. Furthermore, the self-attention mechanism faces rising computational complexity with large context windows. To address these issues, state space models (SSMs)~\cite{ssm} like Mamba~\cite{mamba} have been introduced very recently, showing promising results. The RawBMamba~\cite{rawbmamba} introduces a bidirectional Mamba in anti-spoofing. Nevertheless, the challenge lies in effectively applying Mamba for anti-spoofing tasks with pre-trained models as an alternative to transformer-based frameworks.



This work introduces a new bidirectional Mamba structure referred to as the Dual-Column Bidirectional Mamba (DuaBiMamba) for anti-spoofing. DuaBiMamba contains two separate columns for handling the forward and backward features to have improved ability. The outputs of the two columns are finally merged to capture the local and global feature dependencies. Then we propose a novel Mamba-based structure integrated with the self-supervised model wav2vec 2.0 and refer to it as XLSR-Mamba. The primary studies in this work are carried out on the ASVspoof 2021 database to showcase the effectiveness of the proposed XLSR-Mamba and real-world applicability for the inference time involved. We further evaluate the robustness of our proposed model on a more challenging In-the-Wild dataset and compare it with other SOTA system results.


\begin{figure*}[ht]
\centering
\centerline{\includegraphics[width=0.9\textwidth]{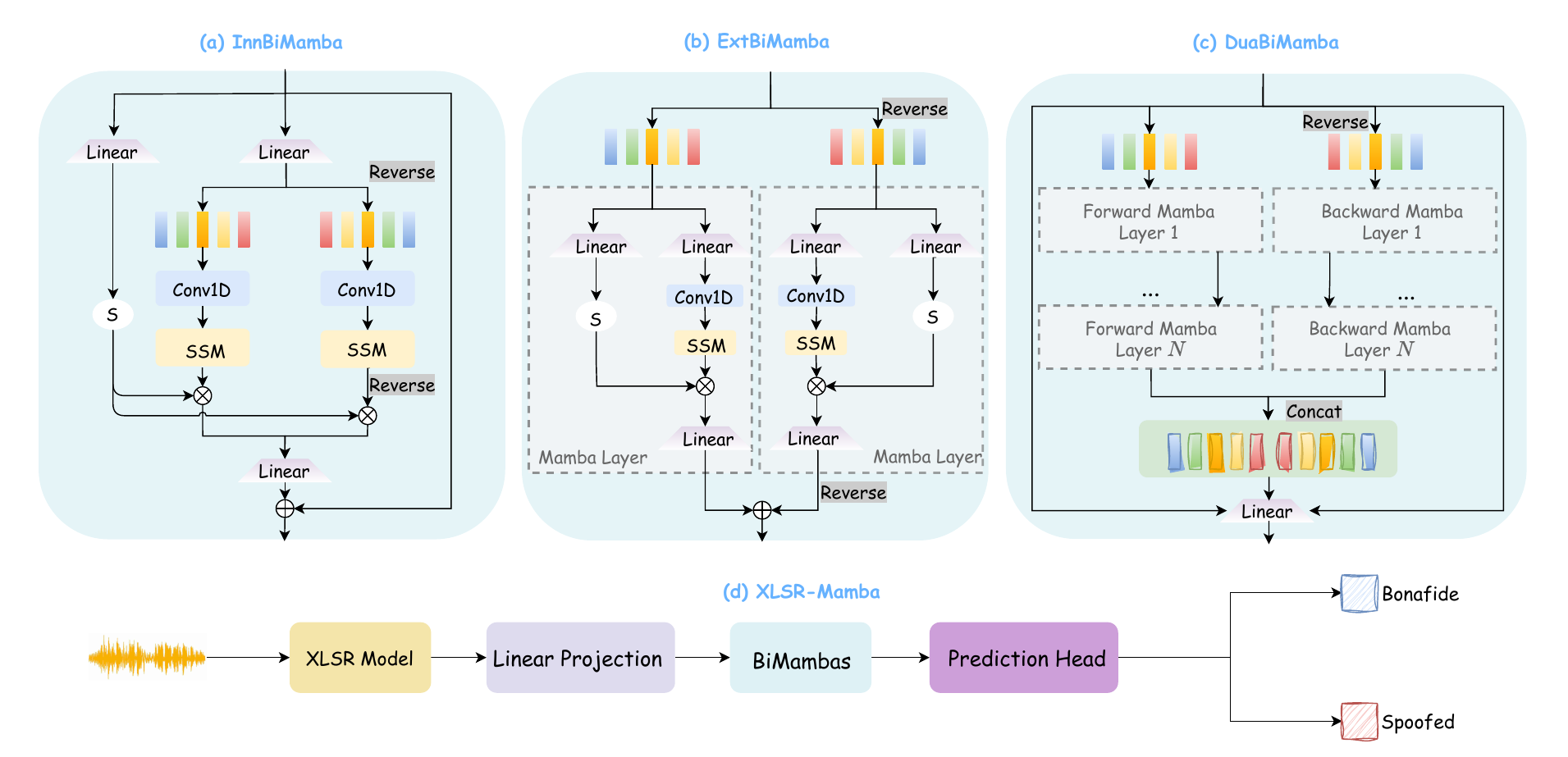}}
\vspace{-3mm}
\caption{Overview of the XLSR-Mamba architecture for anti-spoofing, with three different BiMamba configurations: (a) External Bidirectional Mamba (ExtBiMamba); (b) Inner Bidirectional Mamba (InnBiMamba); and (c) Dual-Column Bidirectional Mamba (DuaBiMamba); (d) XLSR-Mamba model pipeline, including XLS-R feature extraction, linear projection, BiMamba blocks, and prediction head. \blue{The prediction head is implemented as a linear layer that maps features produced by the BiMamba blocks into whether the input speech is bonafide or spoofed.}}
\vspace{-3mm}
\label{fig:overview}
\end{figure*}

\section{Preliminaries}
Structured SSMs~\cite{ssm} efficiently handle long-dependent sequences with low computational and memory demands, making them suitable replacements for transformers or recurrent neural networks (RNNs). Mamba improves upon SSMs by introducing an input-dependent selection mechanism for better information filtering and a hardware-aware algorithm that scales linearly with sequence length, allowing for faster computation~\cite{mamba}. Mamba’s architecture combines SSM blocks with linear layers. It is simple yet achieves the SOTA performance in various long-sequence tasks~\cite{tf}, offering substantial computational efficiency during training and inference.

\blue{At its core, Mamba employs a linear selective SSM. The SSM operates by mapping an input \(x_t\) to an output \(y_t\) through a hidden state \(h_t\) at time \(t\).}
The matrices \(A\), \(B\), and \(C\) are continuous learnable parameters corresponding to the state transition, input, and output processes. \blue{We first discretize the continuous parameters to discrete counterparts \(\tilde{A}\), \(\tilde{B}\). This operation allows the model to process discrete-time audio signals. Then the SSM as the equation below: }

\begingroup
\vspace{-4mm}
\begin{equation}
h_t = \tilde{A} h_{t-1} + \tilde{B} x_t, \quad y_t = C h_t
\end{equation}
\vspace{-6mm}
\endgroup

Due to its linear structure, the entire output sequence \(y\) of length \(\mathcal{L}\) can be expressed as a convolution between the input sequence \(x\) of the same length and a kernel \(\mathcal{K}\). 

\begingroup
\vspace{-6mm}
\blue{\begin{equation}
    \quad y = x \ast \mathcal{K}, {\text {~where~}} \mathcal{K} = (C\tilde{B}, C\tilde{A}\tilde{B}, \ldots, C \tilde{A}^{\mathcal{L}-1}\tilde{B}) 
\end{equation}}
\vspace{-6mm}
\endgroup

\blue{Here, each element in \(\mathcal{K}\) reflects how an input at a given time step is propagated through the state transitions to influence future outputs. This convolution kernel \(\mathcal{K}\) compactly captures the temporal dynamics of the model, enabling efficient processing of long sequences—a key advantage of structured state-space models for anti-spoofing tasks.} However, since the matrices depend on the input \(x_t\) (making them selective), this convolution cannot be computed directly. Instead, it is solved using a parallel scan algorithm. A unidirectional Mamba is an SSM placed between gated linear layers. While these modifications have improved model performance by adding ``selective" features, they do not change the fundamental unidirectional nature of SSMs, similar to RNNs. This unidirectionality is not a problem for training large language models, as many use an autoregressive approach. However, in non-autoregressive speech models for anti-spoofing, a module with non-causal capabilities, like attention mechanisms, is necessary. Thus, it is essential to find an effective solution to address this limitation.

\section{Proposed Method}
\subsection{Bidirectional Mamba}
The original Mamba model performs causal computations in a unidirectional manner, using only historical information. However, bidirectional computations are essential in speech tasks where complete speech is available. Mamba must process information from past and future contexts to capture global dependencies within input features—similar to the multi-head self-attention (MHSA) module. To address this, a previous research~\cite{mambaspeech} introduced two bidirectional strategies for Mamba in speech enhancement and automatic speech recognition: inner bidirectional Mamba (InnBiMamba), and external bidirectional Mamba (ExtBiMamba).

We first explore InnBiMamba shown in Fig.~\ref{fig:overview} (a), adapted from Vision Mamba, for anti-spoofing tasks. In this design, two SSM modules share the same input and output projection layers. One SSM module processes the input in the forward direction, while a time-reversed version of the input is fed to the other module. The output from the backward module is then reversed and combined with the output of forward module, passing through the output projection layer. In contrast, ExtBiMamba as shown in Fig.~\ref{fig:overview} (b) has separate input and output projection layers for the forward and backward Mamba modules. The forward module processes the input, while the backward module handles the reversed input. The outputs are fused through addition, with a residual connection applied around the ExtBiMamba layer.
\subsection{Proposed Dual-Column Bidirectional Mamba}
\blue{Both InnBiMamba and ExtBiMamba have limited adaptability for anti-spoofing tasks. In these models, the forward and backward pathways are merged either through shared input/output projections (InnBiMamba) or simple addition (ExtBiMamba), which restricts their ability to differentiate fine-grained cues—such as unnatural intonation or stress—from broader patterns like a flat emotional tone.}
In this regard, we propose the Dual-Column Bidirectional Mamba (DuaBiMamba) to address these limitations. DuaBiMamba as shown in Fig.~\ref{fig:overview} (c) is designed with two separate Mamba columns: one column processes all forward features, while the other handles all backward features. \blue{Their outputs are then concatenated, which fuses local details (captured via Conv1D layers within each Mamba block) and global dependencies (modeled by the selective state space mechanisms).}
Thereby, this dual-column based Mamba structure captures the complex patterns essential for anti-spoofing.

\subsection{Proposed XLSR-Mamba}
We adopt the XLSR-Conformer~\cite{asvconformer} structure as our baseline architecture to propose the XLSR-Mamba model as shown in Fig.~\ref{fig:overview} (d). This model leverages the pre-trained XLSR, a variant of wav2vec 2.0. With extensive SSL training, the model can extract rich speech representations for spoofing attack detection. The XLSR model consists of two primary components: a convolutional neural network front-end, which converts the 1D raw waveform into a 2D temporal-channel representation, and 24 transformer encoder layers that capture global relationships within the speech signal. The resulting speech representation is a \(T \times 1024\) matrix, where \(T\) denotes temporal length and 1024 is the number of channels. After obtaining the XLSR representation, we project it to a \(D\)-dimensional space. These \((T \times D)\) features are then passed through the BiMamba blocks to learn higher-level representations. The BiMamba composes \blue{\(N\)} Mamba blocks, each containing SSM and Conv1D layers to capture local dependencies in the speech representation. Finally, the output features from the BiMamba model are used to classify the given input speech as bonafide or spoofed.

\section{Experimental Setup}

\subsection{Datasets and Performance Metrics}
To assess the effectiveness and generalizability of the proposed method for anti-spoofing, we conducted experiments on ASVspoof 2021 LA \& DF, and In-the-Wild datasets. All models were trained on the ASVspoof 2019 LA training set~\cite{asvspoof2019}. The ASVspoof 2021 LA evaluation set contains approximately 180,000 utterances, combining bonafide and spoofed speech generated by speech synthesis or voice conversion algorithms~\cite{asvspoof2021}. The ASVspoof 2021 DF evaluation set includes around 600,000 utterances processed with different lossless codecs for media storage.  On the other hand, the In-The-Wild dataset~\cite{inthewild} contains 17.2 hours of spoofed and 20.7 hours of real data. Spoofed data were sourced from 219 public videos promoting speech deepfakes, while bonafide clips from the same speakers came from podcasts and speeches. 

We evaluated model performance in terms of equal error rate (EER) and the minimum tandem detection cost function (min t-DCF)~\cite{tDCF_Tomi_IEEE}, which are the standard evaluation metrics in the ASVspoof Challenges.

\subsection{Implementation Details}
We used the RawBoost~\cite{rawboost} to increase data variability, adding linear and non-linear convolutive noise, impulsive signal-dependent additive noise, and stationary signal-independent additive noise, following the configuration in~\cite{asvconformer}. The models trained for LA track used convolutive and impulsive noise, while that for DF track used stationary noise with random coloration.


During training, audio segments were cropped or concatenated to around 4 seconds (64,600 samples). The Adam optimizer was used with a learning rate of \(10^{-6}\) and a weight decay of \(10^{-4}\) to optimize a weighted cross-entropy loss, with a batch size of 20. Final results were based on an averaged checkpoint of the top-5 validation models, with early stopping after 7 epochs without improvement. We set the embedding size of XLSR to 144 and \blue{\(N\)} of Mamba blocks to 12 as the default configuration. \blue{The Conv1D layers in BiMamba use a kernel size of 3 and the feature dimension \(D\) is set to 256.}
\begin{table}[t]
\caption{Comparison of InnBiMamba, ExtBiMamba, and DuaBiMamba architectures on the ASVspoof 2021 LA and DF evaluation sets. Average means the average EER of both LA and DF sets.}
\label{tab:bimambas}
\centering
\resizebox{0.9\columnwidth}{!}{%
\begin{tabular}{lcccc}
\hline
                                   & \multicolumn{2}{c}{\textbf{ASVspoof 2021 LA}}                & \textbf{ASVspoof 2021 DF}  & \textbf{Average}  \\ \cline{2-5} 
\multirow{-2}{*}{\textbf{Methods}} & EER\%                      & min t-DCF                       & EER\%  & EER\%                      \\ \hline
XLSR+Mamba                     & 1.04                         & 0.215                         & 2.54   & 1.79                      \\
XLSR+InnBiMamba                    & 1.54                         & 0.227                         & \cellcolor[HTML]{C4D5EB}1.82 & 1.68\\
XLSR+ExtBiMamba                     & 0.99                         & 0.212                         & 2.54   & 1.77                      \\
XLSR+DuaBiMamba                    & \cellcolor[HTML]{C4D5EB}0.93 & \cellcolor[HTML]{C4D5EB}0.208 & 1.88  & \cellcolor[HTML]{C4D5EB}1.41                       \\ \hline
\end{tabular}%
}
\vspace{-6mm}
\end{table}

\section{Results and Analysis}
\subsection{DuaBiMamba vs. ExtBiMamba \& InnBiMamba}

Table~\ref{tab:bimambas} shows that DuaBiMamba performs the best on the ASVspoof 2021 LA set, achieving the lowest EER of 0.93\% and min t-DCF. 
\blue{Its dual-column structure, with separate forward and backward Mamba layers, is beneficial for identifying fine-grained artifacts in the spoof data.} However, InnBiMamba achieves the lowest EER of 1.82\% on the DF set.
\blue{This may be attributed to its shared input and output projection layers, which indicate that the architecture is better suited to handling the compression-induced distortions characteristic of DF data.}
ExtBiMamba, while providing separate pathways, may not capture the interaction between forward and backward dependencies as effectively as DuaBiMamba’s concatenation approach. It is also noted that the performance declines without BiMambas as shown in the first row of Table~\ref{tab:bimambas}, indicating its critical role. As DuaBiMamba is better at detecting diverse artifacts of LA and DF data collectively, we propose to use it as the backbone for the XLSR-Mamba model.

\begin{table}[t]
\centering
\caption{Performance comparison with SOTA single systems on the ASVspoof 2021 LA and DF evaluation sets. Here, XLSR-Mamba represents the XLSR model with DuaBiMamba.\\
*The average result from three independent training sessions}
\vspace{-2mm}
\label{tab:sota}
\resizebox{0.9\columnwidth}{!}{%
\begin{tabular}{lccc}
\hline
                                   & \multicolumn{2}{c}{\textbf{2021LA}} & \textbf{2021DF} \\ \cline{2-4} 
\multirow{-2}{*}{\textbf{Systems}} & EER\%               & min t-DCF               & EER\%                     \\ \hline
2021 RawNet2~\cite{rawnet2}                            & 5.31                & 0.310                   & 22.38                     \\
2023 SE-Rawformer~\cite{rawformer}                       & 4.98                & 0.318                   & -                         \\
2024 RawBMamba~\cite{rawbmamba}                          & 3.21                & 0.271                   & 15.85                     \\ \hline
2022 XLSR+AASIST*~\cite{xlsaasist}                      & 1.00                & 0.212                   & 3.69                      \\

2023 XLSR+Conformer~\cite{asvconformer}                    & 0.97                & 0.212                   & 2.58                      \\

2024 XLSR+AASIST2~\cite{aasist2}                      & 1.61                & -                       & 2.77                      \\

2024 WavLM-Large+MFA~\cite{wavlm}                    & 5.08                & -                       & 2.56                      \\
2024 XLSR+Conformer+TCM~\cite{tcmconformer}                 & 1.18                & 0.217                   & 2.25                      \\
\cellcolor[HTML]{C4D5EB}XLSR-Mamba (Proposed)              & \cellcolor[HTML]{C4D5EB}0.93                & \cellcolor[HTML]{C4D5EB}0.208                   & \cellcolor[HTML]{C4D5EB}1.88                      \\ \hline
\end{tabular}%
}
\vspace{-4mm}
\end{table}

\subsection{XLSR-Mamba vs. Other SOTA Models}
The results in Table~\ref{tab:sota} demonstrate that XLSR-Mamba, which integrates the XLSR pretrained model with DuaBiMamba architecture, surpasses other models on the ASVspoof 2021 LA and DF sets. Unlike transformer models, which rely heavily on multi-head self-attention and incur high computational costs to capture context, Mamba achieves efficient processing while preserving rich temporal relationships across sequences. This makes
BiMamba captures long-range feature information more efficiently than Transformer models. Similarly, graph attention networks (e.g., AASIST) are designed to capture complex relationships through graph structures but are not very effective in capturing the nuanced temporal dynamics essential for anti-spoofing, especially when bidirectional information is needed. DuaBiMamba’s architecture is designed to address these temporal dependencies, maintaining a detailed and cohesive representation of speech that strengthens its ability to distinguish between bonafide and spoofed speech. This unique bidirectional structure enables XLSR-Mamba to perform more focused temporal analysis, making it highly effective for anti-spoofing outperforming other SOTA models on the ASVspoof 2021 LA and DF datasets.

\begin{figure}[t]
\centering
\centerline{\includegraphics[width=0.9\linewidth]{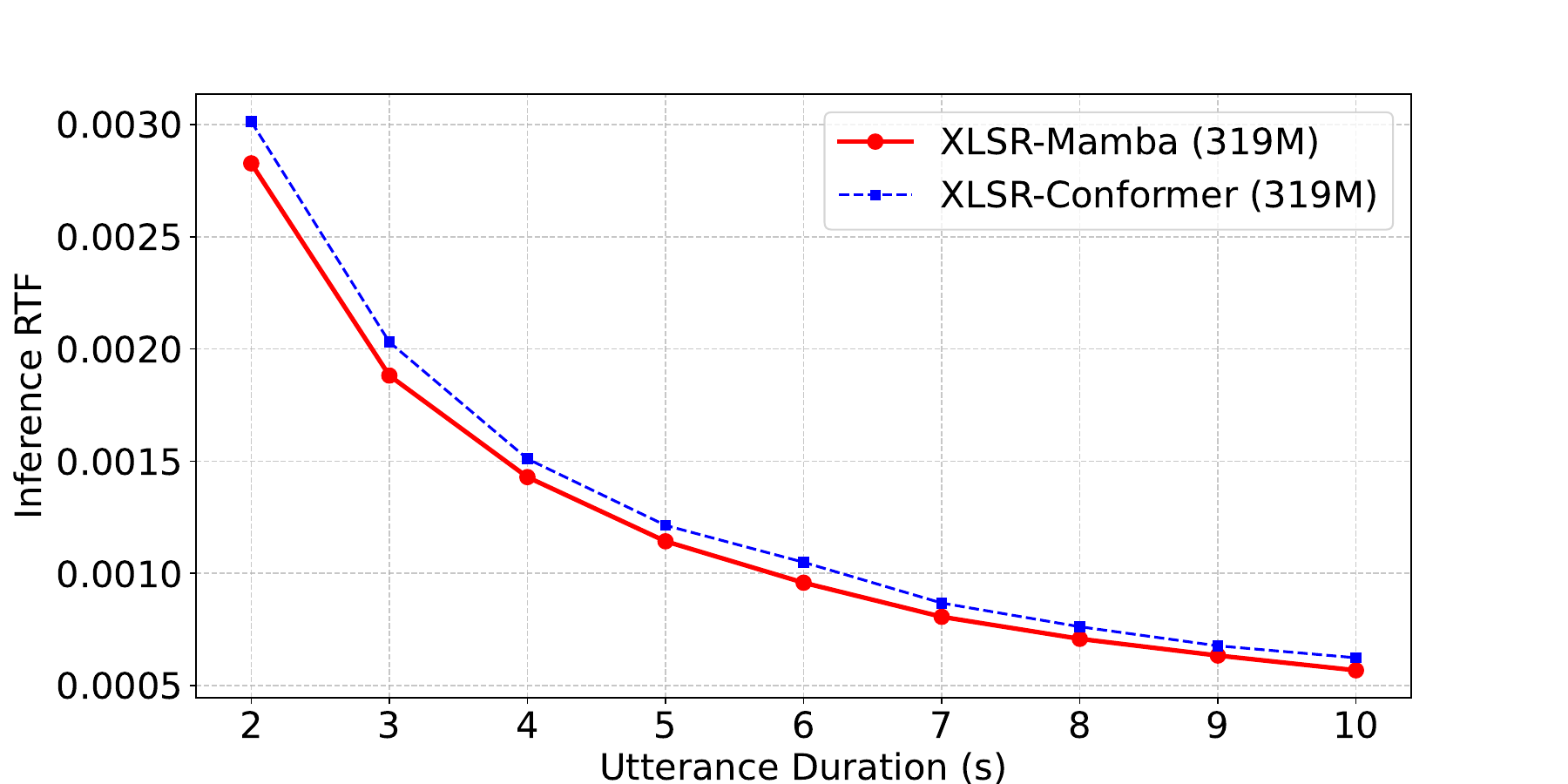}}
\vspace{-3mm}
\caption{Inference speed comparison (real-time factor) between XLSR-Mamba and XLSR-Conformer~\cite{asvconformer} across utterance durations from 2 to 10 seconds.}
\label{fig:curve}
\vspace{-6mm}
\end{figure}

\vspace{-2mm}
\subsection{Efficiency and Effectiveness Analysis of XLSR-Mamba}

Fig.~\ref {fig:curve} demonstrates the inference speed of XLSR-Mamba compared to recent XLSR-Conformer~\cite{asvconformer}, measured using the real-time factor (RTF) on an NVIDIA A5000 GPU, with values averaged over 20 runs. Lower RTF values indicate faster processing relative to the utterance duration. Across varying utterance durations (2 to 10 seconds), XLSR-Mamba consistently achieves a lower RTF than XLSR-Conformer, validating its efficiency. This advantage implies that XLSR-Mamba can process utterances faster, making it more suitable for real-time anti-spoofing applications where efficient processing is critical. 

\begin{figure}[!t]
   \centering
  \begin{minipage}[tbp]{0.45\linewidth}
  \centering
  \includegraphics[width=\linewidth]{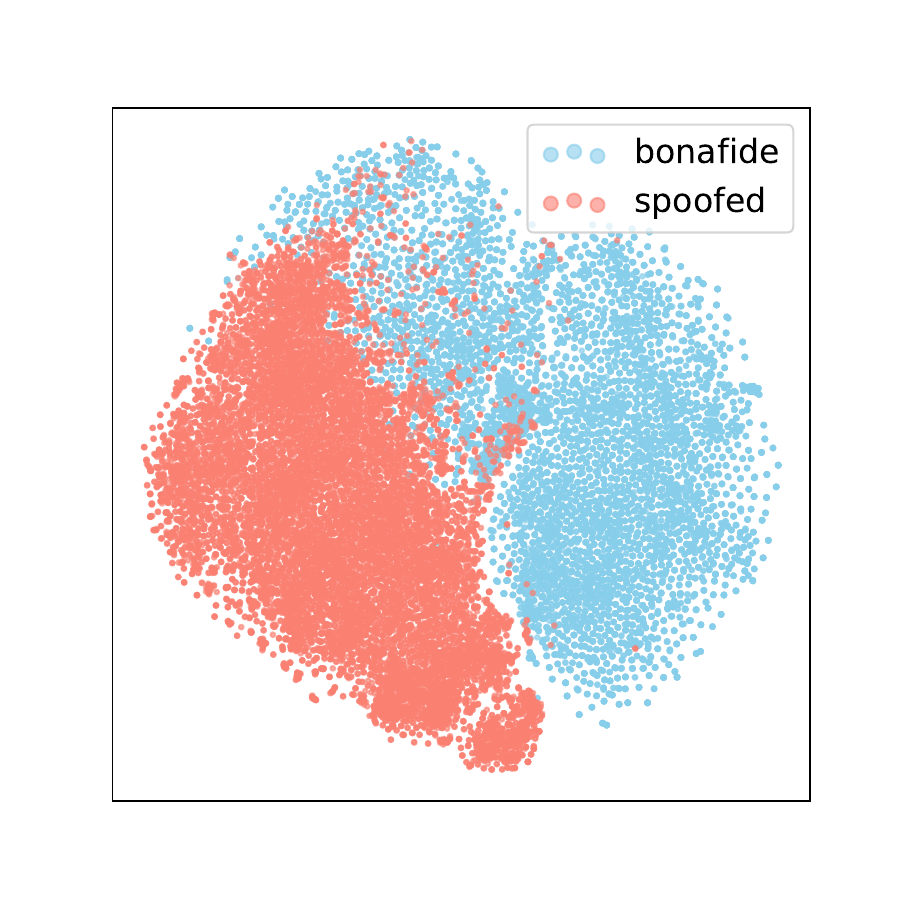}
  \vskip -5mm
  \centerline{\footnotesize{(a) XLSR-Conformer}}
  \end{minipage}
\hfill
  \begin{minipage}[tbp]{0.45\linewidth}
  \centering
\includegraphics[width=\linewidth]{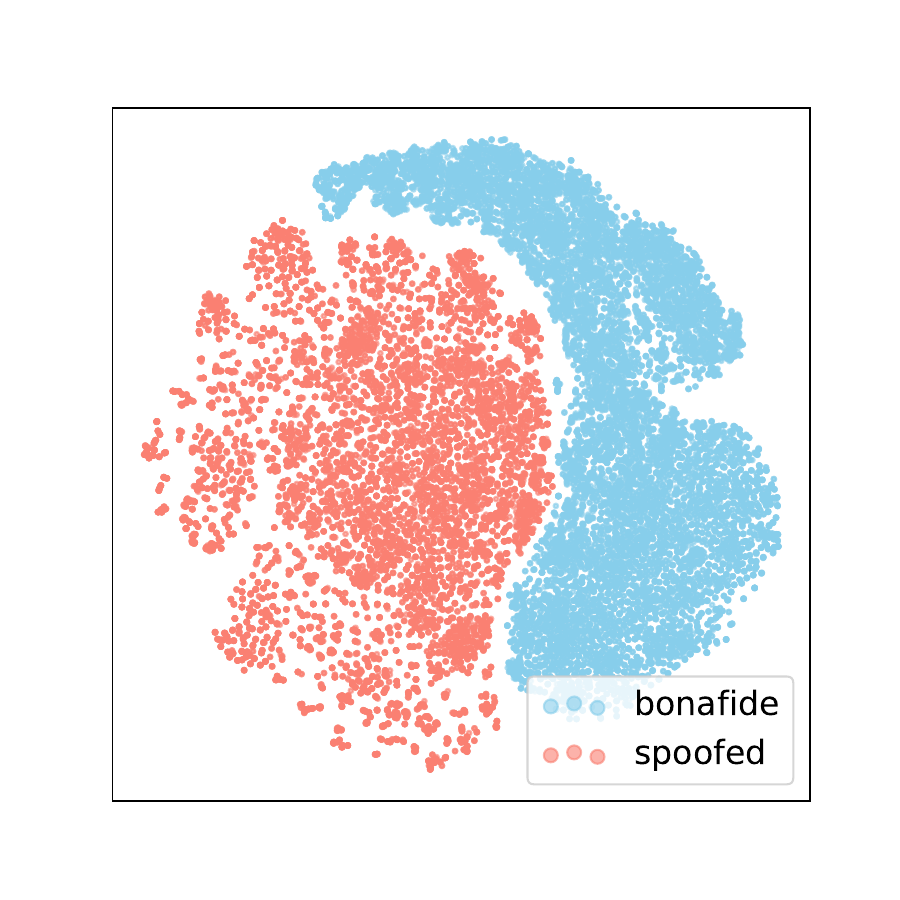}
\vskip -5mm
  \centerline{\footnotesize{(b) XLSR-Mamba}}    
  \end{minipage}
  \vspace{-1mm}
  \caption{The clustering of In-the-Wild test set samples is visualized in 2D t-SNE plots from the model's higher layers.} 
  \label{fig:feat}
  \vspace{-3mm}
\end{figure}

We also extracted final features from the ASVspoof 2021 DF dataset using both the XLSR-Conformer and XLSR-Mamba models, then visualized these features with t-SNE~\cite{tsne}, a nonlinear dimensionality reduction algorithm. Fig.~\ref{fig:feat} show that XLSR-Conformer features display partial overlap between bonafide and spoofed speech clusters, while XLSR-Mamba features have well-separated clusters, suggesting Mamba provides more discriminative features for spoofing attack detection.

\begin{table}[t]
\centering
\caption{Comparison with SOTA single systems on the In-the-Wild dataset. Here, XLSR-Mamba represents the XLSR model with DuaBiMamba architecture.}
\vspace{-2mm}
\label{tab:itw}
\resizebox{\columnwidth}{!}{%
\begin{tabular}{ll}
\hline
                                
\textbf{Systems} & EER\%                \\ \hline
2021 RawNet2~\cite{rawnet2}                       & 33.94 (Reported by~\cite{inthewild})               \\ 2022 XLSR+AASIST~\cite{xlsaasist} & 10.46 (Reported by~\cite{sls})                 \\
2024 XLSR+MoE~\cite{moe}                     & 9.17 (Reported by~\cite{moe}                  \\
2024 XLSR+Conformer~\cite{asvconformer}                     & 8.42 (Our reproduced)                 \\
2024 XLSR+Conformer+TCM~\cite{tcmconformer}     & 7.79 (Evaluated by released checkpoint)                \\
2024 XLSR+SLS~\cite{sls}                     & 7.46 (Reported by~\cite{sls})                \\ 
\rowcolor[HTML]{C4D5EB} 
XLSR-Mamba (Proposed)              & 6.71                 \\ \hline
\end{tabular}%
}
\vspace{-6mm}
\end{table}

\vspace{-4mm}
\subsection{Evaluation on In-the-Wild dataset}
We also evaluate our proposed model with the In-the-Wild dataset which is more challenging and capable of reflecting the generality of models due to similar background noise, emotional tone, and duration across spoofed and genuine instances. Table~\ref{tab:itw} shows that the proposed XLSR-Mamba model achieves the lowest EER of 6.71\%, outperforming other SOTA single systems on this challenging dataset. \blue{The bidirectional state space modeling inherent in XLSR-Mamba allows for better capture of long-range dependencies. }XLSR-Mamba’s structure enhances its sensitivity to long-range timing and rhythm inconsistencies, which are often subtle in deepfake speech and essential for accurately distinguishing between real and spoofed samples in realistic scenarios.

\vspace{-2mm}
\section{Conclusion}
This work presented XLSR-Mamba, a model obtained by combining XLSR pretrained features with a new DuaBiMamba architecture for spoofing attack detection. XLSR-Mamba outperformed other SOTA single systems on the ASVspoof 2021 dataset as well as on the more challenging In-the-Wild dataset. The dual-column bidirectional structure of DuaBiMamba enabled richer temporal features, effectively distinguishing subtle artifacts between genuine and spoofed speech. Additionally, XLSR-Mamba demonstrated faster inference, making it well-suited for real-time applications. These results highlight the promise of Mamba-based architectures over traditional Transformers in voice anti-spoofing.


\balance
\bibliographystyle{IEEEtran}
\bibliography{refs}

\end{document}